\documentclass[preprint]{elsarticle}

\usepackage{lineno,hyperref}
\usepackage{lscape} %usage of landscape
\usepackage{threeparttable} %footnotes in tables
\usepackage{booktabs}
\usepackage{todonotes}
\usepackage{graphicx} %Tablefit to textwidth
\usepackage{placeins} %FloatBarrier
\usepackage{siunitx}
\usepackage{comment}
\usepackage{amssymb}
\usepackage{amsmath}
\graphicspath{{figures/}}
\usepackage{subfig}
\usepackage{comment}   

\usepackage{acronym} 			% abbreviations

\modulolinenumbers[5]

\journal{Renewable and Sustainable Energy Reviews}

\biboptions{sort&compress}
\begin{document}

\begin{frontmatter}

\title{Comparison of power sector models by analyzing the impact of modeling features on optimal capacity expansion}

\author[ISEA,PGS,JARA]{Jonas van Ouwerkerk\corref{correspondingauthor}}
\cortext[correspondingauthor]{Corresponding author}
\ead{jonas.vanouwerkerk@isea.rwth-aachen.de, batteries@isea.rwth-aachen.de}
\author[DLR,STRise]{Hans Christian Gils}
\author[DLR,STRise]{Hedda Gardian}
\author[DIW]{Martin Kittel}
\author[DIW]{Wolf-Peter Schill}
\author[DIW]{Alexander Zerrahn}
\author[FFE]{Alexander Murmann}
\author[RLI]{Jann Launer}
\author[IER,STRise]{Laura Torralba-Díaz}
\author[ISEA,PGS,JARA]{Christian Bußar}

\address[ISEA]{Institute for Power Electronics and Electrical Drives (ISEA), RWTH Aachen University}
\address[PGS]{Institute for Power Generation and Storage Systems (PGS), E.ON ERC, RWTH Aachen University}
\address[JARA]{Jülich Aachen Research Alliance, JARA-Energy}
\address[DLR]{German Aerospace Center (DLR), Institute of Networked Energy Systems, Curiestr. 4, 70563 Stuttgart, Germany}
\address[STRise]{Stuttgart Research Initiative on Integrated Systems Analysis for Energy (STRise), Keplerstraße 7, 70174 Stuttgart, Germany}
\address[DIW]{German Institute for Economic Research (DIW Berlin), Mohrenstraße 58, 10117 Berlin, Germany}
\address[FFE]{Research Center for Energy Economics (FfE), Am Blütenanger 71, 80995 München, Germany}
\address[RLI]{Reiner Lemoine Institute, Rudower Chaussee 12, 12389 Berlin}
\address[IER]{Institute of Energy Economics and Rational Energy Use (IER), University of Stuttgart, Heßbrühlstraße 49a, 70565 Stuttgart, Germany}

\begin{abstract}
The transition towards decarbonized energy systems requires the expansion of renewable and flexibility technologies in power sectors. In a model comparison, we examine the optimal expansion of such technologies with six capacity expansion power system models. The technologies under investigation include base- and peak-load power plants, electricity storage, and transmission. We define four highly simplified and harmonized use cases that focus on the expansion of only one or two specific technologies to isolate their effects on model results. We find that deviating assumptions on limited availability factors of technologies cause technology-specific deviations between optimal capacity expansion in models in almost all use cases. Fixed energy-to-power-ratios of storage can entirely change model optimal expansion outcomes, especially shifting the ratio between short- and long-duration storage. Fixed initial and end storage levels can impact the seasonal use of long-duration storage. Models with a pre-ordered dispatch structure significantly deviate from linear optimization models, as limited foresight and flexibility can lead to higher capacity investments. A simplified net transfer capacity approach underestimates the need for grid infrastructure compared to a more detailed direct current load flow approach. We further find deviations in model results of optimal storage and transmission capacity expansion between regions and link them to variable renewable energy generation and demand characteristics. We expect that the general effects identified in our stylized setting also hold in more detailed model applications, although they may be less visible there. \\

Word count: 6020
%texcount main.tex -inc -incbib -sum -1
\end{abstract}


\end{frontmatter}

\section*{Highlights}
\begin{itemize}
\item We compare six power sector capacity expansion models in a systematic way
\item We use simplified scenarios and fully harmonized input data
\item Four use cases focusing on the expansion of power plants, storage, and transmission
\item Modeling of plant availability, storage designs, and transmission drives deviations
\item Pre-defined dispatch order causes large deviations from deterministic optimization
\end{itemize}

\newpage

\section*{List of abbreviations}
% Abbreviations
\begin{acronym}
 \acro{DCLF}{direct current load flow}
 \acro{E2P}{energy to power}
 \acro{EES}{electric energy storage}
 \acro{LP}{linear programming}
 \acro{NTC}{net transfer capacity}
 \acro{PV}{photovoltaics}
 \acro{TPP}{thermal power plants}
 \acro{TYNDP}{Ten-Year Network Development Plan}
 \acro{VRE}{variable renewable energy}
\end{acronym}

\newpage

\section{Introduction}\label{sec:introduction}

\subsection{Background and motivation}\label{sec:background_motivation}

The 2020 European Climate Law as part of the European Green Deal sets emission reduction targets of 55\% until 2030 and targets climate neutrality for 2050 \cite{EU21}. However, only 18\% of the European gross energy consumption were covered by renewable energy sources in 2018 \cite{VincentJacquesleSeigneur.2019}. The use of renewable energy is a main strategy for decarbonizing not only the power sector, but also the heat provision and mobility sectors via sector coupling. Therefore, decarbonizing the power sector is one of the main challenges when combating climate change by expanding renewable generation capacities. Variable renewable energy sources such as solar PV and wind power generally have low costs and high expansion potentials, yet they require additional system flexibility \cite{Lopez2020} beyond what base load power plants are able to provide. Short- and long-duration storage as well as the transmission grid facilitate a temporal and spatial smoothing of \ac{VRE}, complemented with flexible \ac{TPP} \cite{Schill2020,Schlachtberger2017}.

Capacity expansion models are used to investigate possible pathways for the transition of the power sector. These models are per definition simplified representation of the real world, in most cases minimizing total system costs. The abstraction from reality leads to a variety of different modeling approaches that may lead to diverging results, e.g. concerning the optimal expansion of generation and flexibility technologies energy systems with \ac{VRE}.

\subsection{State of research}\label{ssec:state_of_research}
There are only few studies in the literature that compare the outcomes of capacity expansion models in a structured way. Most of it focuses on the theoretical comparison of certain aspects of existing models. Gacitua et al.~\cite{Gacitua2018} discuss policy instruments of 21 capacity expansion models for the power sector. Dagoumas et al.~\cite{DAGOUMAS20191573}, categorize capacity expansion models as optimization, equilibrium and alternative models, and evaluate advantages and disadvantages. Only a few studies compare capacity expansion scenarios for different models. Gils et al.~\cite{Gils2019} compare four high-resolution power sector models with three scenario variations for the German energy system in 2050. Siala et al.~\cite{siala2020} compare six power market models, focusing on model type, planning horizon and temporal as well as spatial resolution. Both studies highlight the importance of harmonizing input data for understanding result deviations in comparative scenario analyses. However, the complexity of the conducted scenarios impede the identification of drivers of individual differences. To gain a deeper understanding of the underlying effects and drivers of capacity expansion model outcomes from different models, simplified scenarios can be useful \cite{Gils2021b}.

\subsection{Contribution of this paper}\label{ssec:contribution}

This paper contributes to the energy modeling literature by exploring the drivers for differences in the outcomes of capacity expansion models. To this end, technology expansion outcomes of six power sector models are compared. In a novel approach, highly simplified use cases separately examine the expansion of simple combinations of power generation, storage, or transmission technologies. This approach addresses the challenge of data harmonization \cite{siala2020}, substantially reduces complexity regarding model outcomes, and enables the association of outcome differences with model specifics. The selected scope of model approaches and technology modeling of the six contributing power sector models aims at covering a wide range of typical model features. This increases the significance and usefulness of results for a wider group of model users and developers. Insights may also support policy makers when interpreting model-based policy recommendations. The broad scope and simplicity of the defined use cases allow us to trace back the influence of the hourly \ac{VRE} generation and demand profiles on the model results.

\section{Materials and Methods}\label{sec:material_methods}

Section \ref{ssec:set_up_model_comparison} and \ref{ssec:data_characteristics} introduce the set-up and input data of the model comparison. Section \ref{ssec:contributing_models} briefly presents contributing models and their properties. Section \ref{ssec:model_differences} highlights key differences between these models relevant to the discussion of results. Finally, Section \ref{ssec:the_procedure} describes the procedure the model comparison and result analysis.

\subsection{Scenario set-up and data collection}\label{ssec:set_up_model_comparison}

\begin{figure}[htp]
    \centering
    \includegraphics[width=8cm]{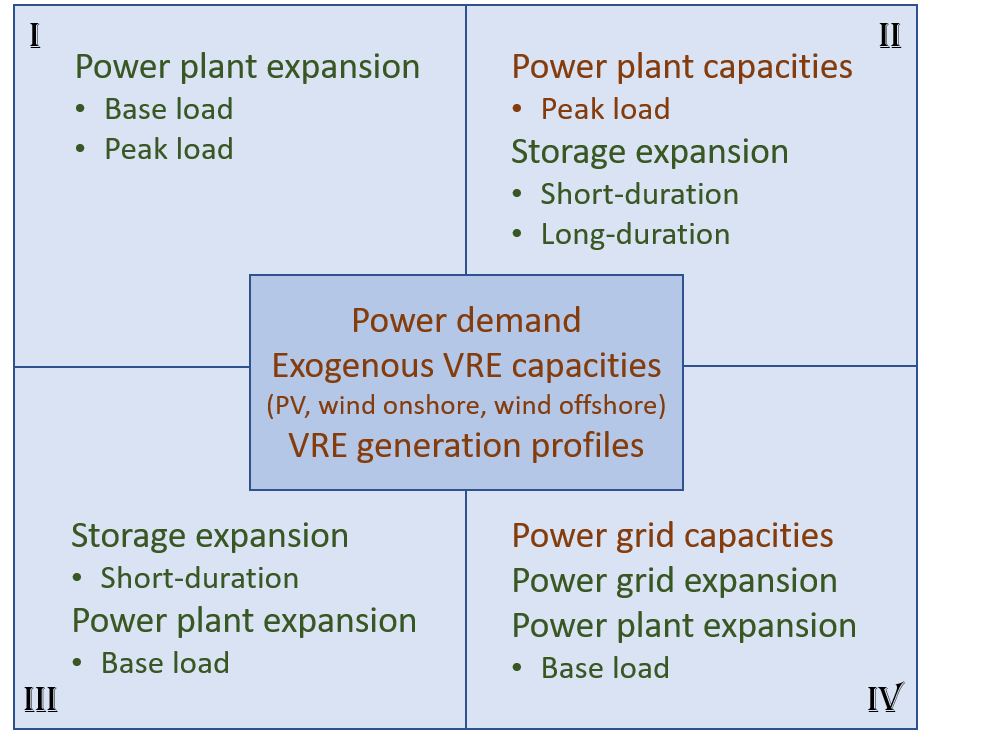}
    \caption{Use cases considered in the model comparison. Endogenous capacity optimization is indicated by "expansion" (green), exogenously defined capacities by "capacities" (orange). The capacities in the center are identical for all use cases. }
    \label{fig:use_cases}
\end{figure}

The model comparison comprises four abstract use cases of a future Central European energy system with regional demand and \ac{VRE} generation potential profiles. The geographical scope includes the countries of Austria, Belgium, Czech Republic, Denmark, France, Germany, Italy, Luxembourg, the Netherlands, Poland, and Switzerland. All of the simplified systems consist of the same hourly power demand time series and hourly generation potentials of three \ac{VRE} technologies including \ac{PV}, wind onshore, and wind offshore. Further, pre-installed and expandable technology capacities are individually defined for all use cases. Figure \ref{fig:use_cases} gives an overview of the available technologies and endogenous or exogenous capacity choices in the different use cases. All use cases are optimized in an hourly temporal resolution for a full year.

Use case I focuses on the capacity optimization of dispatchable power plants. We include nuclear power plants as an example for a base load generation technology with high investment costs and low variable costs, and gas turbines as an sample peak load technology with lower investment costs but higher variable costs. Use case II investigates the capacity optimization of \ac{EES}, with lithium-ion batteries as short-duration storage option and hydrogen caverns with electrolyzers and hydrogen turbines for long-duration storage. To avoid a lack of supply, peak load (gas) turbines with a capacity matching the region-specific residual load peak are also included in use case II. The competition between dispatchable generation and storage is addressed in use case III, where the capacities of base load power plants and short-duration \ac{EES} are optimized. Finally, use case IV evaluates the interaction in the capacity expansion of power transmission lines and base load power plants. Power exchange across model regions is only possible in use case IV. Existing and currently planned transmission capacities are considered according to ENTSO-E's \ac{TYNDP} \cite{ENTSOE.}. Further, an endogenous expansion of those capacities is allowed. Across all use cases, input data is completely harmonized, including techno-economic parameters, \ac{VRE} capacities, annual demand as well as the corresponding time series. The full input data set is provided in this issue (Gils et al.~\cite{Gils2021b}). The highly simplified design of the use cases enables linking the differences in models outcomes to model specifics, and does not intend to realistically represent the future European energy system.

\subsection{Data characteristics}\label{ssec:data_characteristics}

The profiles for the \ac{VRE} generation potentials and power demand profiles show an individual pattern per region. To highlight such regional characteristics we define a PV-to-demand-ratio ($\rho_{PV,demand,r}$) (\ref{eq:PV_d_ratio}), Wind-to-demand-ratio ($\rho_{Wind,demand,r}$) (\ref{eq:Wind_d_ratio}), summer-demand-share ($\rho_{summer,winter,r}$) (\ref{eq:summer_d_share}), and winter-demand-share ($\rho_{winter,summer,r}$) (\ref{eq:winter_d_share}). In this context, summer is defined from 21st of March until 20th of September and winter from 21st of September until 20th of March. 

\begin{equation} 
\label{eq:PV_d_ratio}
\begin{split}
    \rho_{PV,demand,r} = \frac{1}{T} \sum_{t=1}^T \frac{C_{PV,r} \cdot c_{PV,r}(t)}{d_{r}(t)} \quad \forall r \in \text{\{Regions\}} \\
\end{split}
\end{equation}

\begin{equation} 
\label{eq:Wind_d_ratio}
\begin{split}
    \rho_{wind,demand,r} = \frac{1}{T} \sum_{t=1}^T \frac{C_{Wind,r} \cdot c_{Wind,r}(t)}{d_{r}(t)} \quad \forall r \in \text{\{Regions\}} \\
\end{split}
\end{equation}

\begin{equation} 
\label{eq:summer_d_share}
\begin{split}
    \rho_{summer,demand,r} = \frac{\sum_{t=21.Mar}^{t=20.Sep} d_{r}(t)}{\sum_{t=01.Jan}^{t=31.Dec} d_{r}(t)} \quad \forall r \in \text{\{Regions\}} \\
\end{split}
\end{equation}

\begin{equation} 
\label{eq:winter_d_share}
\begin{split}
    \rho_{winter,demand,r} = 1-\rho_{summer,demand,r} \quad \forall r \in \text{\{Regions\}} \\
\end{split}
\end{equation}

where $C$ is the power plant capacity, $c$ is the capacity factor for \ac{VRE} units, and $d$ is the demand.

By using a large number of regions with different characteristics, we aim at strengthening the data foundation and robustness of our analysis. Further, this gives us the opportunity to analyse the effects of region-specific settings on the expansion of technologies. 

\begin{figure}[htbp]
\centering
\subfloat[Wind-to-demand-ratio.\label{fig:wind-load-ratio}]
    {\includegraphics[width=0.47\textwidth]{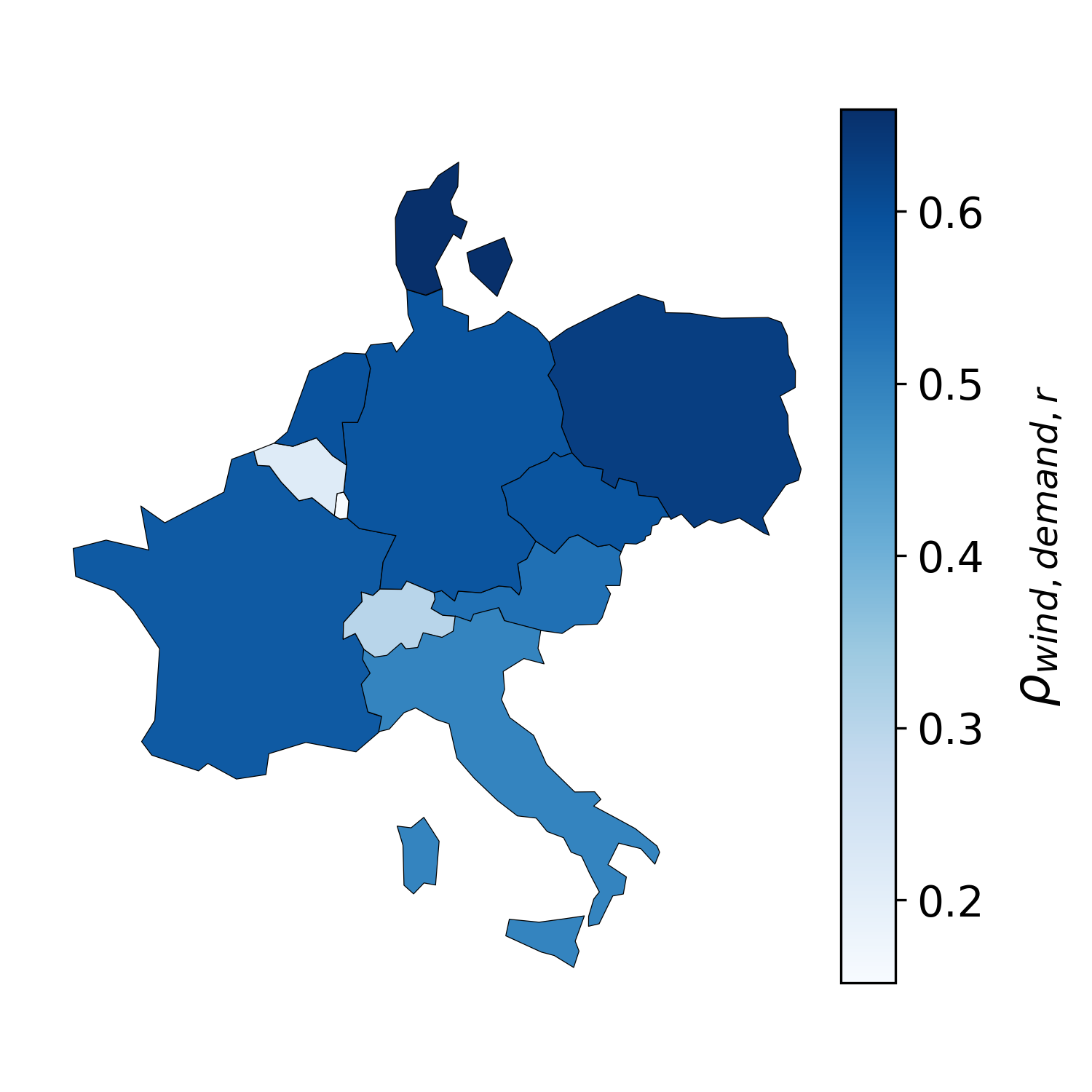}}
\qquad
\subfloat[PV-to-demand-ratio.\label{fig:pv-wind-ratio}]
    {\includegraphics[width=0.47\textwidth]{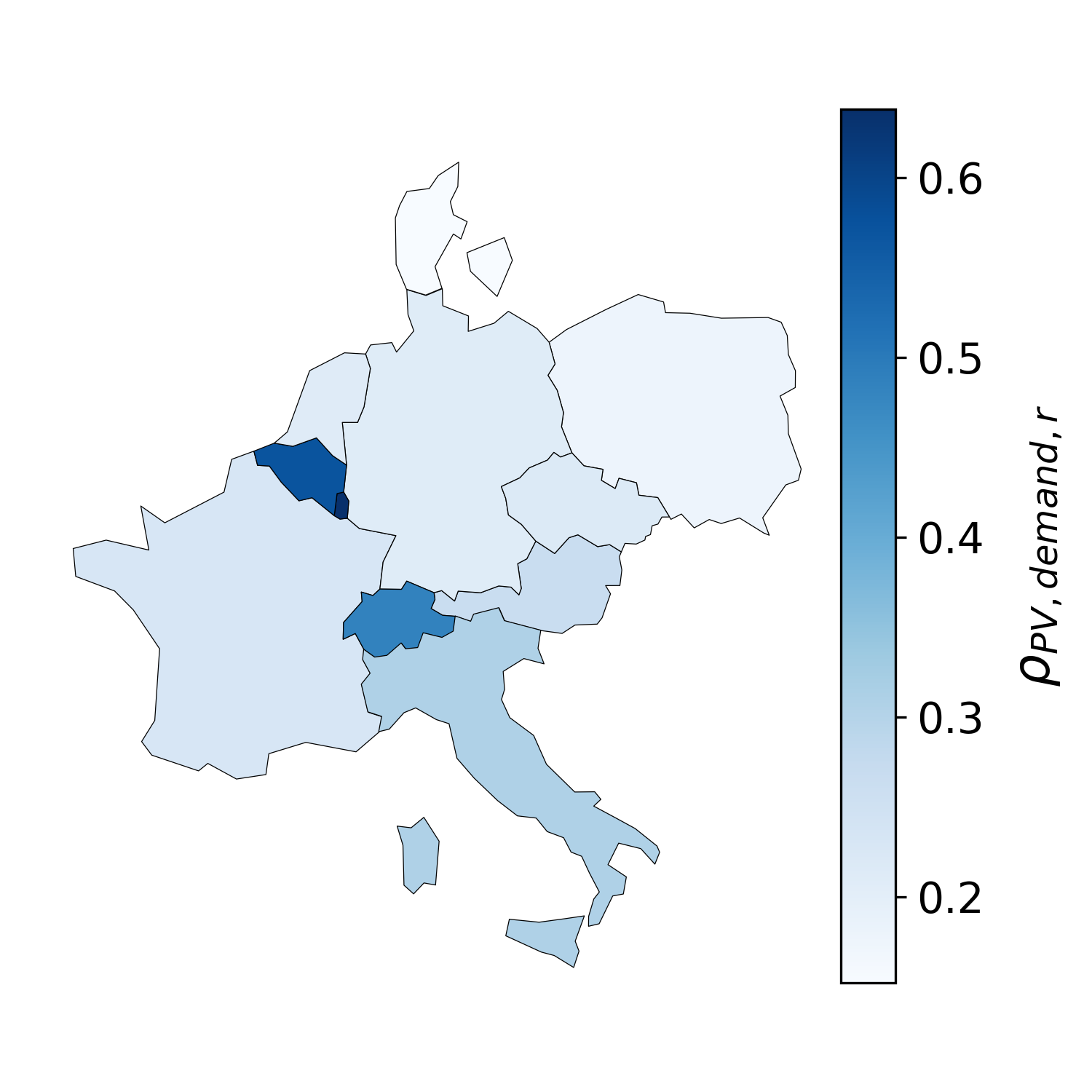}}\\
\subfloat[Summer-demand-ratio.\label{fig:summer-demand-ratio}]
    {\includegraphics[width=0.47\textwidth]{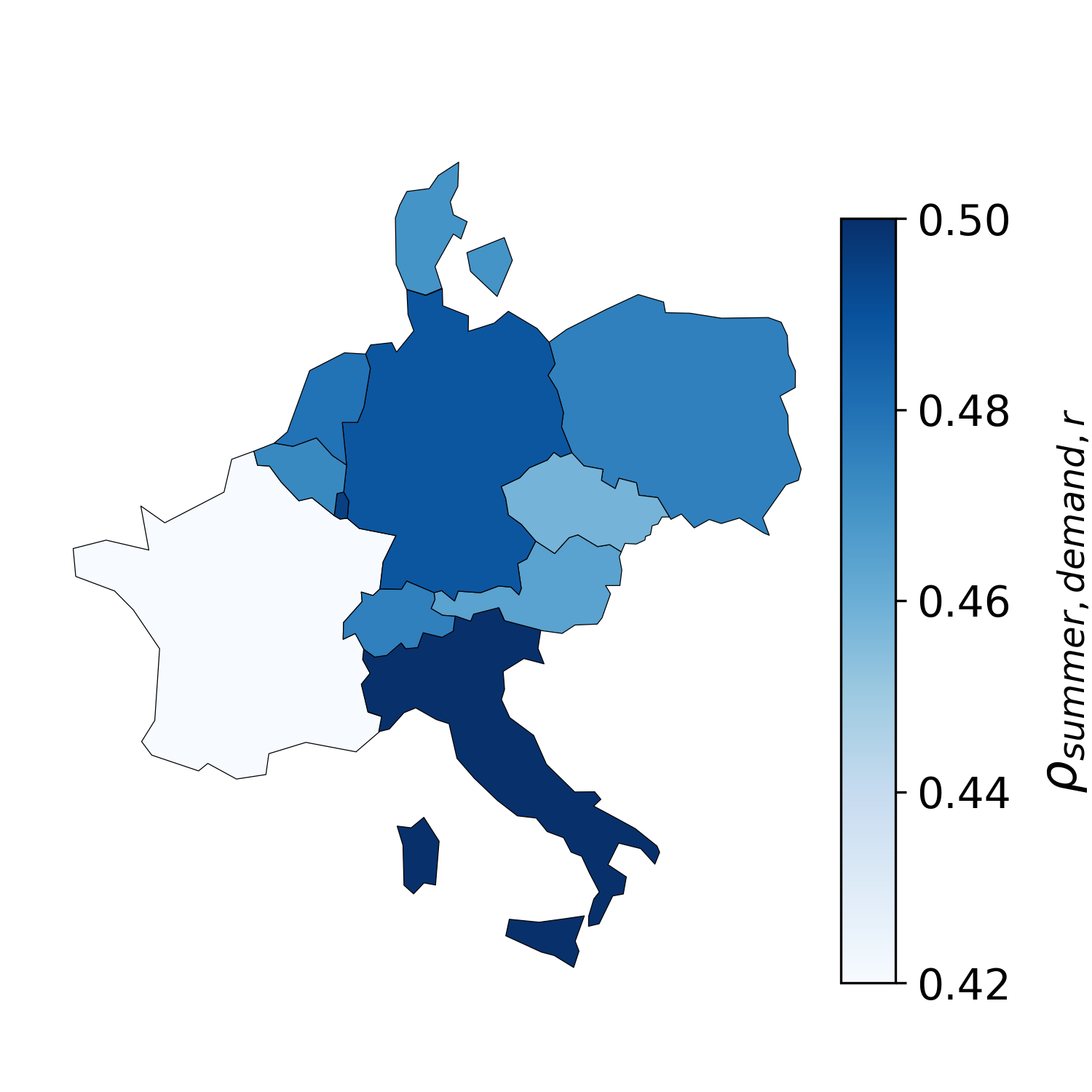}}\\
\caption{Regional \ac{VRE} and demand characteristics.}%
\label{fig:capDiffRegion}%
\end{figure}

Figure \ref{fig:capDiffRegion} illustrates the characteristics (1-3) for all regions, which vary strongly across the considered regions. Poland and Denmark have the highest wind-demand-ratios, followed by Germany, Czech Republic, France, and the Netherlands. In contrast, the solar-demand-ratio is especially high in Luxembourg, Belgium, Switzerland, and Italy. While in France energy consumption in summer is 13\% lower than in winter, demand is similar in winter and summer in Italy. 

It is important to emphasize that the regional characteristics used here are highly stylized, as they for example neglect hydro power. They should thus not be used for deriving real-world policy conclusions. Yet, their simplified structure allows for meaningful comparisons and insights in the context of this model comparison exercise.

\subsection{Contributing models}\label{ssec:contributing_models}

In this paper, six power sector models with maintainers from different research institutions across Germany contribute to the model comparison (Table \ref{tab:model_overview}). All of them feature an hourly time resolution, a multi-regional setting with grid interconnection, and an optimized capacity expansion of generation, balancing, and transmission technologies. They differ in modeling language, programming approach and the level of foresight. \mbox{GENESYS-2} follows a different approach than the other \ac{LP} models by building on a population-based heuristic to find the optimal solution. Additionally, the underlying dispatch structure sets a fixed rule-based dispatch hierarchy for all generation and storage technologies. For this reason, the model has no foresight beyond the actual hour of operation. Another feature of the dispatch structure is that it drives local use of energy and short-distance energy distribution. More information on these models can be found in this issue (Gils et al.~\cite{Gils2021b}).

\begin{table}[htp]
\setlength{\tabcolsep}{0.8mm}
\caption{Overview of the six contributing models and their properties in the versions used for this model comparison.}
\begin{scriptsize}
\label{tab:model_overview}
\begin{tabular}{lcccccc}
\toprule
 & \multicolumn{1}{l}{\textbf{DIETER}} & \textbf{E2M2} & \textbf{GENESYS-2} & \textbf{ISAaR} & \textbf{oemof} & \textbf{REMix} \\
\midrule
Programming language & GAMS & GAMS & C++ & \begin{tabular}[c]{@{}l@{}}MATLAB, \\ PostgreSQL \end{tabular} & Python & GAMS \\ \\
Modeling approach & LP & LP & \begin{tabular}[c]{@{}c@{}}population- \\based \\ heuristic\end{tabular} & LP & LP & LP \\ \\
%Solver & CPLEX & CPLEX & CMA-ES &  CPLEX & CPLEX & CPLEX & CBC & CPLEX & fmincon \\
Deterministic & yes & yes & optional & yes & yes & yes \\ \\
Foresight in hours & 8760 & 8760 & none & 8760 & 8760 & 8760 \\ \\
Documentation & \cite{zerrahn_2017,schill2018,gaete2020} & \cite{Sun2013,Torralba2020} & \cite{Bussar.2019,Siemonsmeier.2018} & \cite{ISAaR-Dok,Pellinger2016,Boeing2020} & \cite{HILPERT201816, KRIEN2020100028,oemof_solph_documentation}  & \cite{Gils2017,Gils2017a}  \\
\bottomrule
\end{tabular}
\end{scriptsize}
\end{table}

\subsection{Key modeling differences}\label{ssec:model_differences}

The contributing models differ with respect to technology representation and model features (Table \ref{tab:model_tech_overview}). Typical features, however, coincide across most models and only differ for some of the models. This allows for isolating the impact of different implementations on model outcomes. For instance, in contrast to all other models, GENESYS-2, oemof, and ISAaR do not consider a limited power plant and storage availability, which requires less generation and storage capacity for supplying peak load. The same implications apply to power plant flexibility which is only considered in DIETER and REMix by modeling simple load change costs. Regarding electricity storage, DIETER requires initial and final storage levels to equal 50 \% of the endogenous energy storage capacity, which potentially affects the seasonal operation of long-duration storage. In E2M2, both the \ac{E2P} ratio and the ratio between charging and discharging capacity of storage units are exogenously fixed, resulting in a more limited flexibility when expanding such technologies. Furthermore,the models have fundamentally different approaches for power transmission modeling. While REMix uses a detailed \ac{DCLF} approach, all other models use a simplified transport model based on \ac{NTC} for grid representation. The \ac{NTC}-based approach used in GENESYS-2, however, differs from the other models. It is embedded in the pre-defined dispatch order, ensuring that transmission is only possible when there is a regional surplus from \ac{VRE} or a shortage in local generation. In case a region requires import or export via transmission, the transmission model favours exchange with neighbouring regions over more distant regions.

Note that many of these characteristics solely apply to the particular model version used for this comparison exercise, and do not represent the full capabilities of the models. Many of these simplifications are, among others, a result of the data harmonization process.

\begin{table}[htp]
\setlength{\tabcolsep}{0.8mm}
\caption{Overview of technology modeling details and features that apply to the model version used for this comparison.}
\label{tab:model_tech_overview}
%\resizebox{\textwidth}{!}{%
\begin{scriptsize}
\begin{tabular}{@{}lllllll@{}}
\toprule
Feature & \textbf{DIETER} & \textbf{E2M2} & \textbf{GENESYS-2} & \textbf{ISAaR} & \textbf{oemof} & \textbf{REMix} \\ \midrule

\begin{tabular}[c]{@{}l@{}}Limited \\ technology \\ availability\end{tabular} & \begin{tabular}[c]{@{}l@{}}constant \\ factor\end{tabular} & \begin{tabular}[c]{@{}l@{}}constant \\ factor\end{tabular} & \begin{tabular}[c]{@{}l@{}}not \\ considered\end{tabular} & \begin{tabular}[c]{@{}l@{}}not \\ considered\end{tabular} & \begin{tabular}[c]{@{}l@{}}not \\ considered\end{tabular} & \begin{tabular}[c]{@{}l@{}}constant \\ factor\end{tabular} \\ \\

\begin{tabular}[c]{@{}l@{}}Power plant \\ flexibility \end{tabular} & \begin{tabular}[c]{@{}l@{}}simple load \\ change costs\end{tabular} & \begin{tabular}[c]{@{}l@{}}no load \\ change costs\end{tabular} & \begin{tabular}[c]{@{}l@{}}no load \\ change costs\end{tabular} & \begin{tabular}[c]{@{}l@{}}no load \\ change costs\end{tabular} & \begin{tabular}[c]{@{}l@{}}no load \\ change costs\end{tabular} & \begin{tabular}[c]{@{}l@{}}simple load \\ change costs\end{tabular} \\ \\

\begin{tabular}[c]{@{}l@{}}Storage \\ levels\end{tabular} & \begin{tabular}[c]{@{}l@{}}start: \SI{50}{\percent}, \\ end: \SI{50}{\percent}\end{tabular} & \begin{tabular}[c]{@{}l@{}}optim., \\ equal \end{tabular} & \begin{tabular}[c]{@{}l@{}}start: \SI{0}{\percent}, \\ end: optim.\end{tabular} & \begin{tabular}[c]{@{}l@{}}start: \SI{0}{\percent}, \\ end: optim.\end{tabular} & \begin{tabular}[c]{@{}l@{}}optim., \\ equal \end{tabular} & \begin{tabular}[c]{@{}l@{}}optim., \\ equal \end{tabular} \\ \\

\begin{tabular}[c]{@{}l@{}}Storage \\ \ac{E2P} ratio \end{tabular} & variable & fixed & variable & variable & variable & variable \\ \\

\begin{tabular}[c]{@{}l@{}}Charging to \\ discharging \\ cap. ratio \\ long-duration \\ storage \end{tabular} & variable & \begin{tabular}[c]{@{}l@{}}fixed \end{tabular} & variable & variable & variable & variable \\ \\

\begin{tabular}[c]{@{}l@{}}Power \\ transmission \end{tabular} & \begin{tabular}[c]{@{}l@{}}with \\expansion\\ \ac{NTC}-based \end{tabular} & \begin{tabular}[c]{@{}l@{}} \end{tabular} & \begin{tabular}[c]{@{}l@{}}with \\expansion\\ \ac{NTC}-based \end{tabular} &  & \begin{tabular}[c]{@{}l@{}}without \\expansion\\ \ac{NTC}-based \end{tabular} & \begin{tabular}[c]{@{}l@{}}with \\expansion\\ \ac{DCLF}-based \end{tabular} \\

%\begin{tabular}[c]{@{}l@{}}Power \\ transmission \\ losses \end{tabular} & ? &  & ? & ? & ? & yes \\

\bottomrule
\end{tabular}
\end{scriptsize}
\end{table}

\subsection{Procedure of the model comparison}\label{ssec:the_procedure}

Not every model participates in each use case (Table \ref{tab:participationMapping}). This is because some models cannot use the harmonized input data, or some model features do not allow for modeling certain use cases. It is important to note that oemof participates in use case IV but without modeling capacity expansion of transmission lines. However, due to the brownfield approach of use case IV, including pre-installed transmission capacities, oemof functions as a benchmark for the other models that allow for transmission expansion.

\begin{table}[htp]
\centering
\begin{scriptsize}
\caption{Overview of model participation in the defined use cases.}
\label{tab:participationMapping}
\begin{tabular}{l|cccc}
\toprule
&\textbf{Use case I} & \textbf{Use case II} & \textbf{Use case III} & \textbf{Use case IV} \\
\midrule
DIETER & \checkmark & \checkmark & \checkmark & \checkmark \\
E2M2 & \checkmark & \checkmark & \checkmark &  \\
GENESYS-2 & \checkmark & \checkmark & \checkmark & \checkmark \\
ISAaR &  &  & \checkmark &  \\
oemof & \checkmark & \checkmark & \checkmark & (\checkmark) \\
REMix & \checkmark & \checkmark & \checkmark & \checkmark \\
\bottomrule
\end{tabular}
\end{scriptsize}
\end{table}

The model comparison follows four steps that at the same time represent the flow of data. First, the input data of the defined use cases is provided to all modelers in a standardized data format. All model maintainers convert this unified data-set into their model specific input format by implementing model specific interfaces. Then, each use case is optimized individually by all participating models. Subsequently, relevant model outcomes are converted into a standardized format, which is designed to allow for an easy and automated comparison. They comprise annual and hourly values reflecting endogenous capacity expansion, generation, transmission, storage use, \ac{VRE} curtailment, or costs. We develop and apply a visualization tool to create standardized plots. This procedure allows for a profound analysis of result deviations, and provides the foundation to associate them with model differences.
%(Gardian et al.~\cite{todo}) 

The comparison of model results requires a harmonized definition of system costs ($K_{system}$), which are minimized in all models. The equation in (\ref{eq:cost}) defines the calculation of total system costs. They consist of annualized investment costs $K_{Invest,annuity}$ as well as fixed ($K_{OPEX,fix}$) and variable ($K_{OPEX,variable}$) operational expenditures. The latter are the sum of fuel costs, emission costs, costs of uncovered load (slack), and power plant specific variable costs. The annualized investment costs and fixed operational expenditures solely include costs for endogenous capacity expansion and are summarized as expansion cost $K_{exp}$.

\begin{equation} 
\label{eq:cost}
\begin{aligned}
    K_{system} & = K_{Invest,annuity} + K_{OPEX,fix} + K_{OPEX,variable} \\
               & = K_{exp} + K_{OPEX,variable}
\end{aligned}
\end{equation}

\section{Results and discussion}\label{sec:results_discussion}

\subsection{Expansion cost over all use cases}\label{ssec:systemcosts}

\begin{figure}[htp]
    \centering
    \includegraphics[width=6cm]{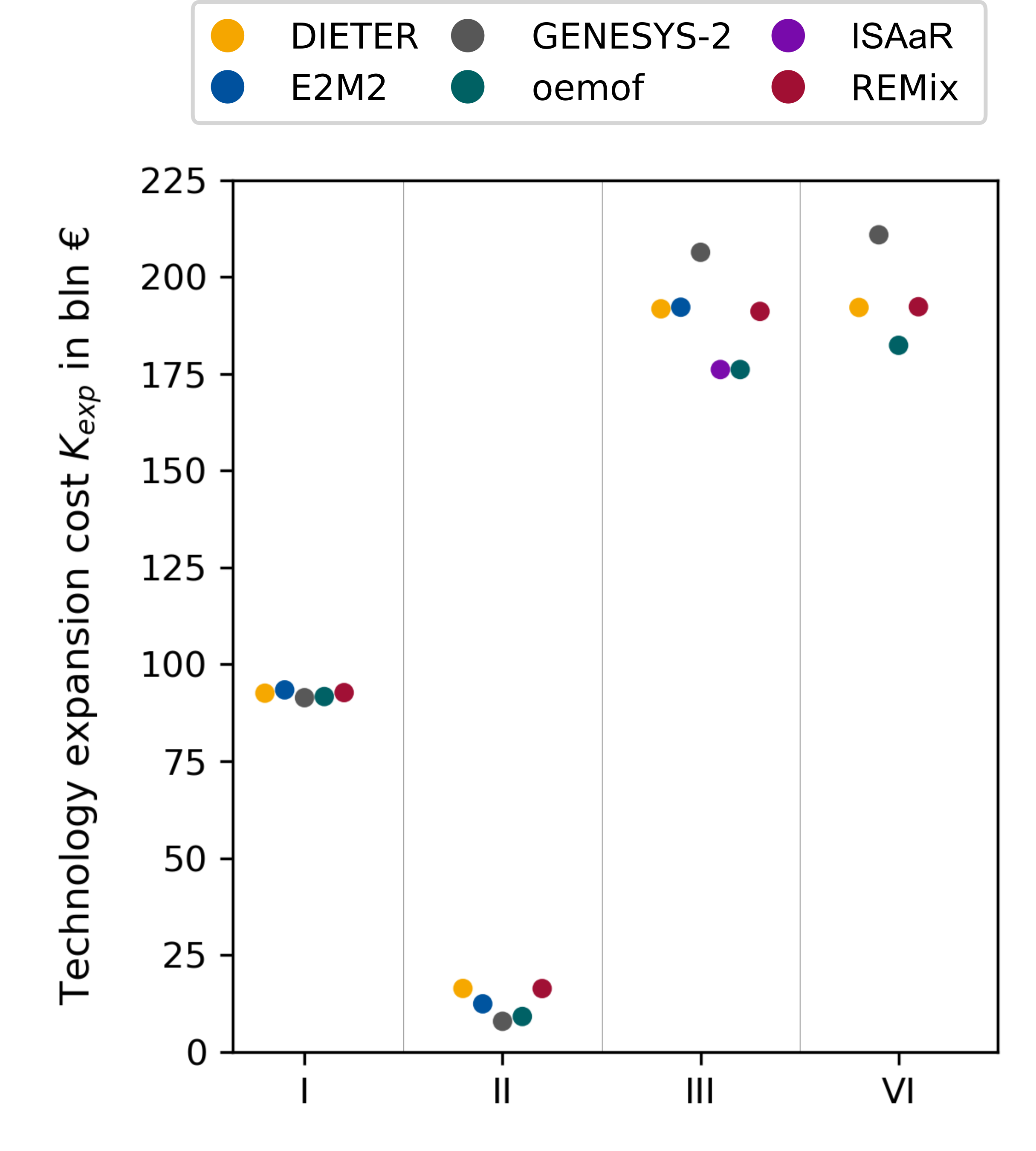}
    \caption{Expansion cost $K_{exp}$ for all use cases I-IV and participating models}
    \label{fig:costComparison}
\end{figure}

The comparison of expansion costs $K_{exp}$ can be used to identify significant result variations originating from different modeling approaches. With the design of the simplified use cases, it is possible to isolate the effects of the expansion of different technologies. In this way, for every expansion technology we can analyse the resulting expansion cost deviations. Figure \ref{fig:costComparison} illustrates the expansion costs for all use cases and models. While expansion costs hardly vary across all models for use case I, they deviate more substantially for use case II-IV. This is driven by relatively similar expansion decisions for use cases I, and substantially different expansion decisions for use cases II-IV. In contrast to use case I and II, that expand capacities of the same technology group (base/peak load power plants or storage), in use cases III and IV entirely different technologies (base/peak load power plants, storage, or transmission) compete against each other for expansion, causing the highest deviations.

In use cases II-IV, expansion cost in oemof are the lowest of all models. One driver for this behaviour is that oemof usually contains less restrictions from features (Table \ref{tab:model_tech_overview}) than other \ac{LP} models. In contrast, GENESYS-2 clearly exceeds the expansion cost of the other models in use cases III and IV. This implicates that the substantially different modeling approach in GENESYS-2 has an overall influence on expansion decisions.

\subsection{Use case I: Expansion of thermal power plants}\label{ssec:expThermal}

\begin{figure}[htp]
    \centering
    \includegraphics[width=\textwidth]{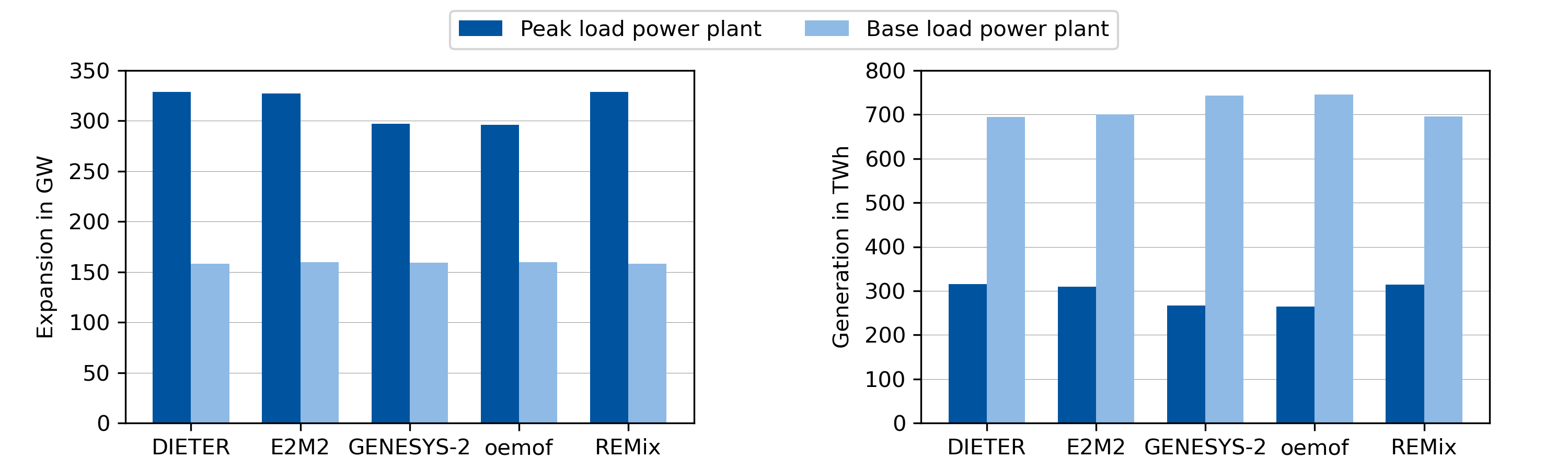}
    \caption{Capacity expansion and generation of thermal power plants across all regions}
    \label{fig:capExpThermal2a}
\end{figure}

Figure \ref{fig:capExpThermal2a} shows the capacity expansion and power generation for base load (nuclear) and peak load (gas) generators accumulated across all regions. 

We find two clusters of model results with similar optimal capacity expansion (Figure \ref{fig:capExpThermal2a}). The first cluster includes DIETER, E2M2, and REMix, the second GENESYS-2 and oemof. Optimal peak load (gas) generation capacity in the first cluster exceeds the second cluster in average by 10\%, while optimal base load (nuclear) capacity is almost identical for both clusters with a maximum deviation of 1\%. These effects are the opposite for generation (Figure \ref{fig:capExpThermal2a}). Models in the first cluster show 6\% less generation from base load, and 18\% more generation from peak load power plants. This is because models use a different availability ratio of peak load over base load power plants. Models in the first cluster apply a constant limitation of available generation capacity to account for planned and unplanned outages. The assumed availability is 94.8\% for peak load (gas) power plants and 91.2\% for base load (nuclear) power plants, i.e.,~the availability ratio is roughly 1.04. Models in the second cluster assume perfect availability with an availability ratio of 1.00. Consequently, each unit of installed base load capacity can generate more electricity than peak load capacity in models of the second cluster compared to those of the first. Additionally, an availability ratio of 1.00 requires less expansion to supply peak loads. Since the specific investment costs of peak load (gas) power plants are substantially lower than those of base load (nuclear) plants, the capacity effect is more pronounced for optimal peak load (gas) capacity (Figure~\ref{fig:capExpThermal2a}). In contrast, the generation effect is rather similar for both technologies. This is because the difference in variable costs of both technologies is lower compared to investment costs. The analysed effects, however, only cause minor variations in expansion cost between the models (Figure \ref{fig:costComparison}). 

Overall, the results indicate that even a subtle model feature such as generation technology availability may cause significant distortions of model outcomes. At the same time, the expansion values indicate that mainly peak load (gas) power plants are affected by this, despite a higher availability. In contrast, the dispatch approach of GENESYS-2 has no impact on model outcomes in this stylized setting as the results are very similar to oemof. This is because the dispatch hierarchy of GENESYS-2 only depends on the marginal costs, and this is what matters in this use case without storage.

\subsection{Use case II: Expansion of storage technologies}\label{ssec:expStorage}

\begin{figure}[htp]
    \centering
    \includegraphics[width=\textwidth]{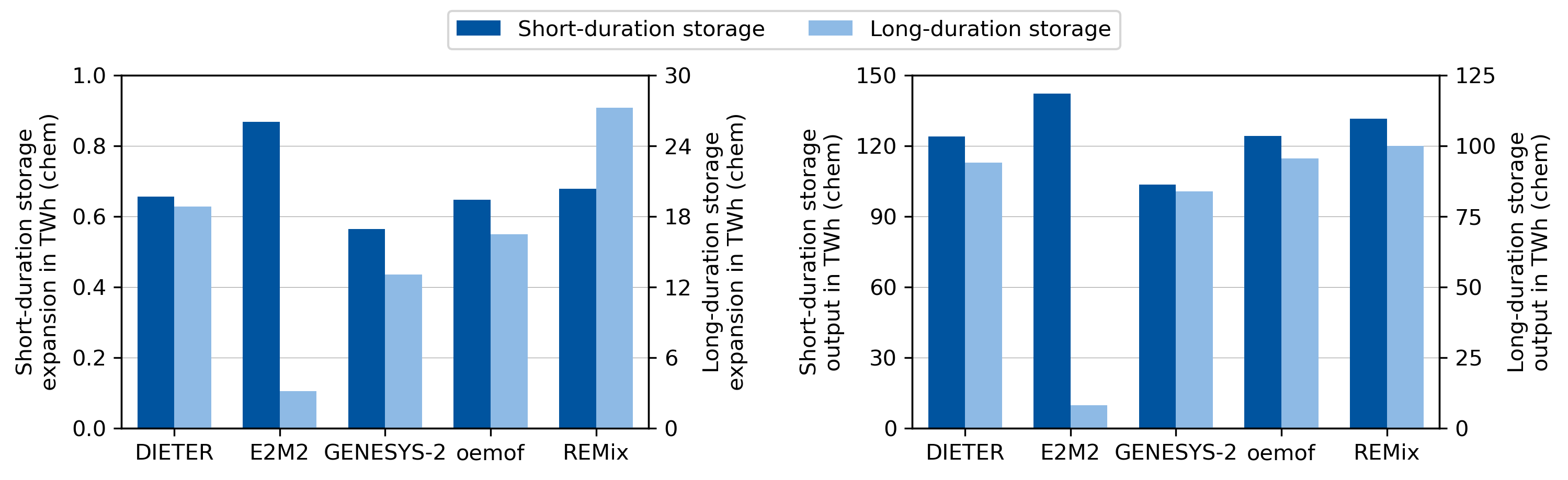}
    \caption{Capacity expansion and discharged energy of short-duration (battery) and long-duration (cavern) storage across all regions.}
    \label{fig:capExpBat2b}
\end{figure}

Figure \ref{fig:capExpBat2b} shows optimal capacity expansion and discharged energy for short-duration (battery) storage and long-duration (cavern) storage for use case~II. %hier unteren Absatz anschließen. Ich verschiebe den nicht, da dabei manchmal Kommentare verloren gehen

In contrast to all other models, E2M2 almost exclusively invests into short-duration storage. The preference towards short-duration storage in E2M2 originates from the fixed \ac{E2P} ratio (4 h for batteries, 400 h for long-duration (cavern) storage) and the fixed ratio between charging and discharging capacity for long-duration storage units. For the charging unit of long-duration (cavern) storage, the exogenous \ac{E2P} ratio in E2M2 is much higher than the endogenous \ac{E2P} ratios determined by the other models, with a maximum of 197 h. For the discharging unit, the optimized values range between 236 h and 1573 h with an average of 620 h. This leads to long-duration (cavern) storage being a more expensive technology to expand in E2M2 than in the other models, which causes a shift from long-duration storage to short-duration storage.

In REMix, we observe an opposite trend with disproportionate expansion of long-duration storage. When applying the storage level limitation to the available storage capacity, the discharging efficiency is taken into account in REMix, as the former is related to the stored energy form and the latter to electrical energy. This leads to lower storage investment costs and higher actually usable storage capacity compared to the other models. Yet, this especially applies to long-duration (cavern) storage which has a relatively low discharging efficiency, and only has a limited effect on short-duration (battery) storage. As a result, the specific costs of storage in REMix are reduced, which in turn results in greater expansion.

For GENESYS-2, the investments into both short- and long-duration storage are generally lower compared to other models. Due to the limited foresight used in GENESYS-2, less energy can be used and stored from \ac{VRE}. For this reason, the capacity expansion of electricity storage is generally lower in  models with a pre-determined dispatch order such as GENESYS-2. Generation from peak load (gas) power plants compensates for the resulting decrease in storage discharge in peak periods. 

\begin{figure}[htp]
    \centering
    \includegraphics[width=10cm]{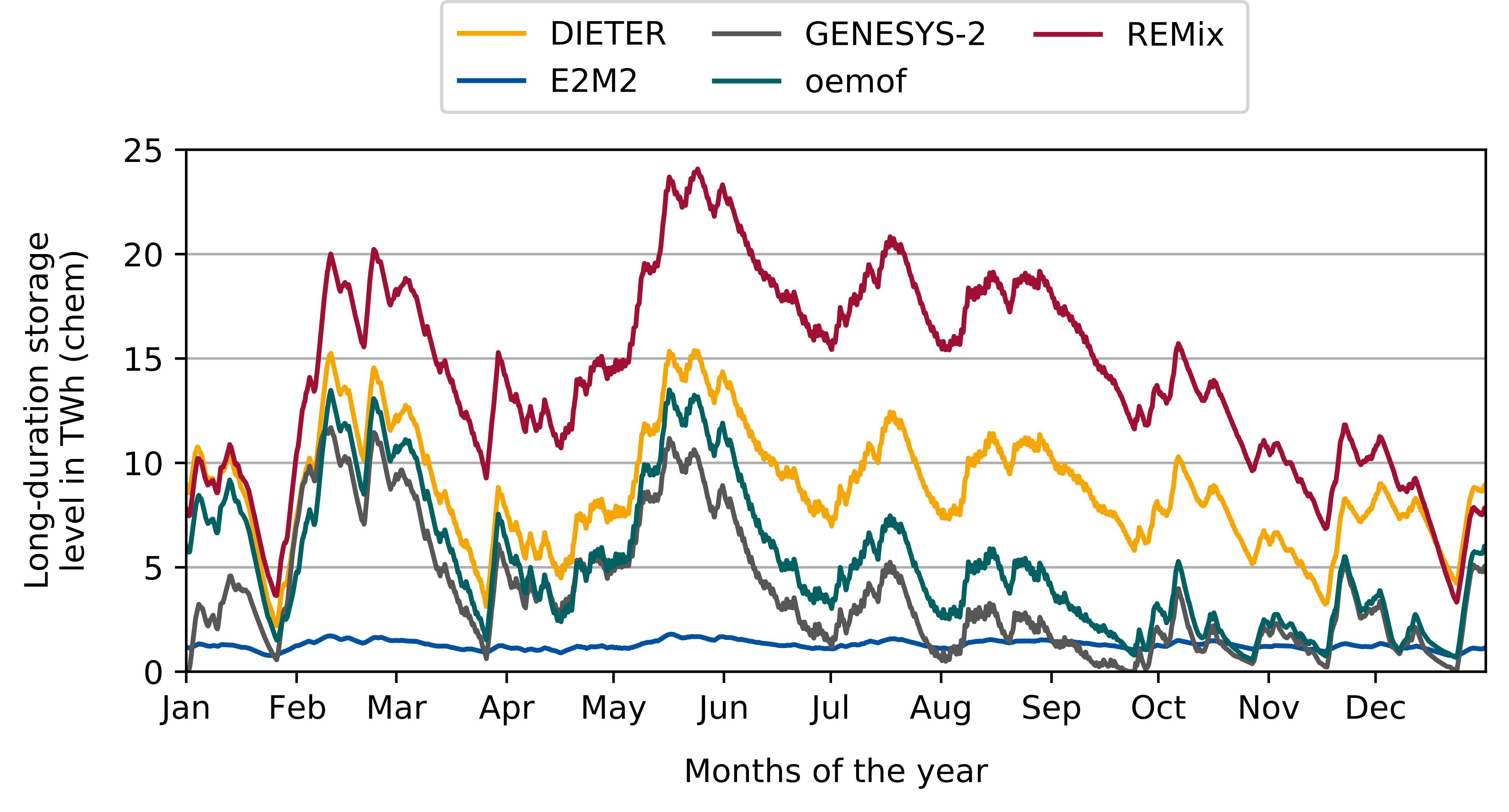}
    \caption{Cavern storage level summed up for all regions.}
    \label{fig:storLevel2}
\end{figure}

The only models with very similar storage expansion are DIETER and oemof. For short-duration (battery) storage, expansion and discharge both show deviations of less than 2\% between the two models. In terms of long-duration (cavern) storage investments, however, DIETER exceeds oemof and shows the second highest expansion of all models. An analysis of the long-duration storage level over the full modeling period, accumulated for all regions, helps to understanding this difference (Figure \ref{fig:storLevel2}). While at the start of the modeling period the (absolute) storage filling level of DIETER and oemof is similar, it deviates over the course of the year. The reason is that the storage level is exogenously set to 50\% of the (optimized) capacity at the beginning and the end of the modeled period in DIETER. Therefore, the storage level cannot drop as much as in the other models towards the end of the year. In contrast, oemof finds the average optimal storage level at the start and end of the simulation at below 50~\%. Additionally, DIETER, in contrast to oemof, assumes a limited availability of long-duration (cavern) storage at 95 \%, which increases the required capacity expansion of this technology.

In a next step, we also examine the distribution of deviations in storage expansion results on a regional level. The scenario set-up contains 11 regions and thus allows for a detailed analysis of results for individual regions. Section \ref{ssec:data_characteristics} highlights important characteristics of the input time series data in each region. Such characteristics include maximum available \ac{VRE} generation shares and seasonal demand variations. These temporal variations have an influence on the optimal use of storage. To quantify the deviating regional patterns, we define the deviation $\sigma_{dev,r,s}$ within the results of one region ($r$), and within the expansion $C_{exp}$ of either short- or long-duration storage ($s$), according to the following measure:

\begin{equation} 
\label{eq:deviation}
\begin{split}
    \sigma_{dev,r,s} = \frac{\max C_{exp,r,s,i} - \min C_{exp,r,s,i}}{\max C_{exp,r,s,i}} \quad \forall i \in \text{\{M\}}, \forall s \in \text{\{S\}}, \forall r \in \text{\{R\}} \\
    \text{with R} \hat{=} \text{\{Regions\}, } \text{M} \hat{=} \text{\{Models\}, } \text{S} \hat{=} \text{\{Storage technologies\}}
    % \text{with R = \{AT,BE,CH,CZ,DE,DK,FR,IT,LU,NL,PL\}} \\
    % \text{with M = \{DIETER,E2M2,GENESYS-2,oemof,ISAaR,REMix\}} \\
    % \text{with S = \{short-term storage, long-term storage\}}
\end{split}
\end{equation}

\begin{figure}[htbp]
\centering
\subfloat[Short-duration (battery) storage.\label{fig:diff_batt_regions}]
    {\includegraphics[width=0.47\textwidth]{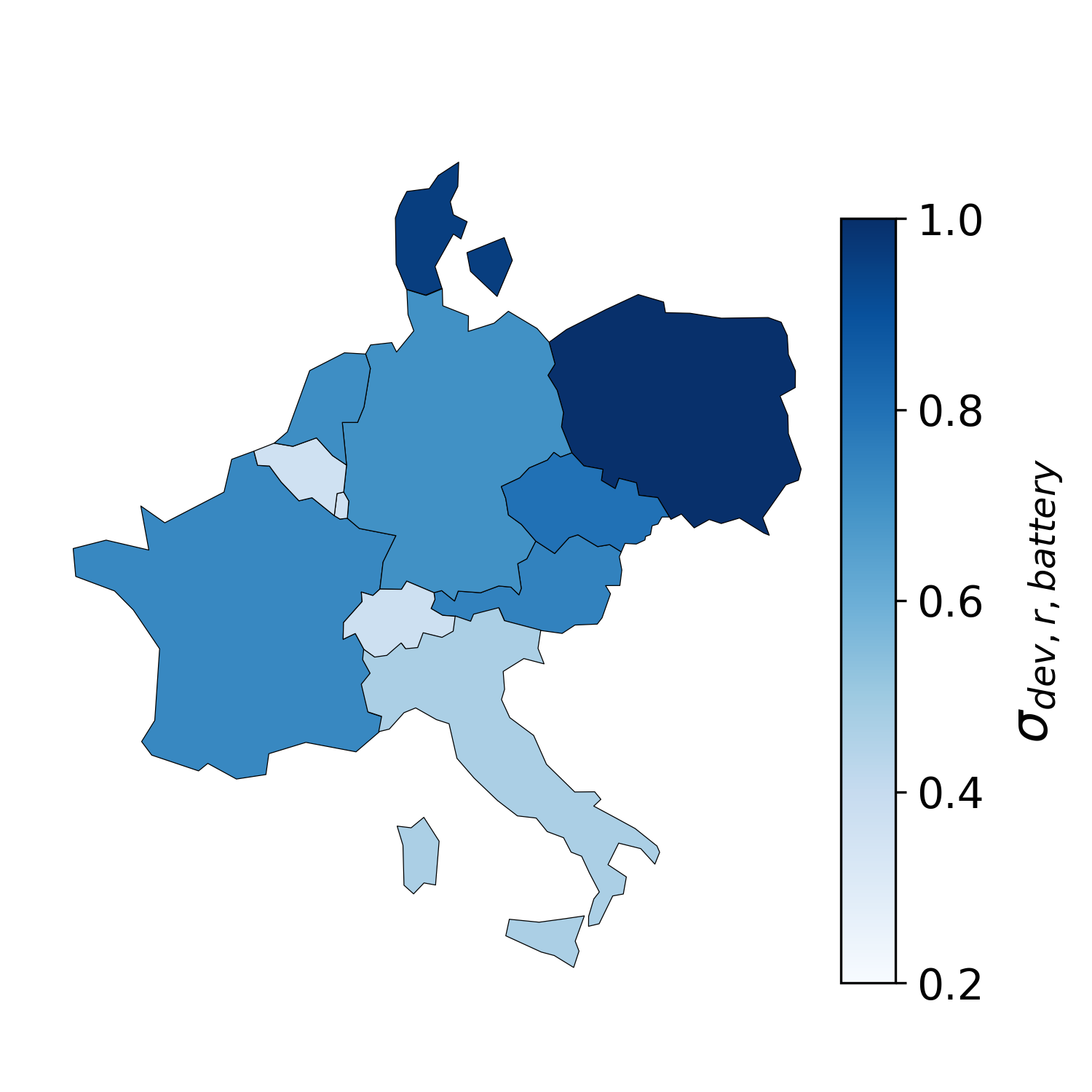}}
\qquad
\subfloat[Long-duration (cavern) storage.\label{fig:diff_carvern_regions}]
    {\includegraphics[width=0.47\textwidth]{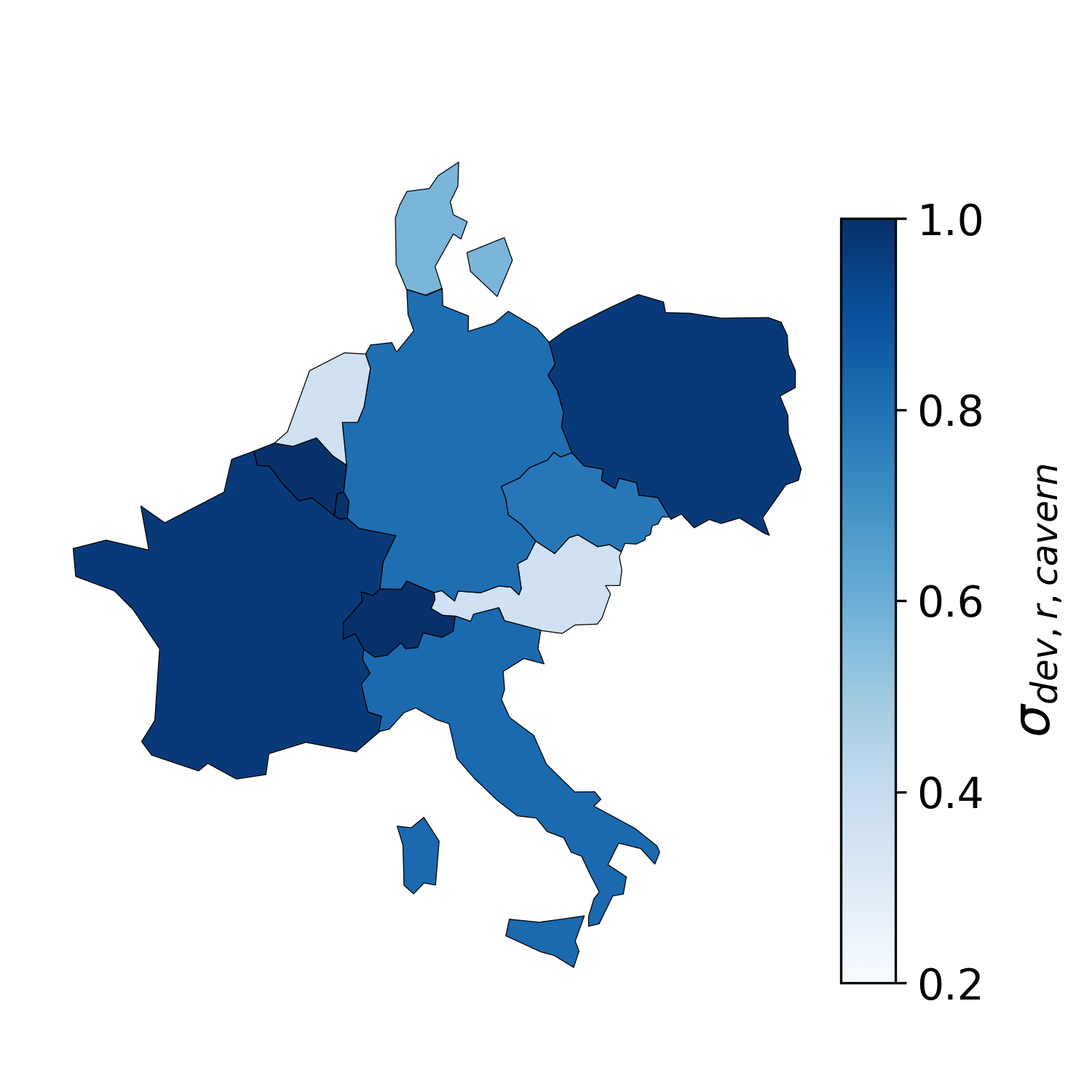}}\\
\caption{Max deviation of optimal storage capacity expansion. A value of zero implies that capacities are identical across all models, whereas the deviations between the models increase for higher values.}%
\label{fig:capDiffStorage}%
\end{figure}

Figure \ref{fig:capDiffStorage} shows the deviation of optimal short-duration (battery) storage capacity and long-duration (cavern) storage capacity for all considered regions.

Optimal short-duration (battery) storage sizing largely coincides between models in Belgium, Italy, Luxembourg, and Switzerland. In Denmark and Poland, two countries with a high share of wind generation (Figure~\ref{fig:capDiffRegion}), the deviation is more than 80\% (Figure \ref{fig:diff_batt_regions}). More generally, results show that deviations of optimal short-duration (battery) storage expansion increase with relatively high generation from wind power. 
In regions with a high \ac{PV} share, short-duration (battery) storage capacity expansion takes place in all models. There are only small deviations between models, and storage is generally large relative to the peak load. In regions with a higher importance of wind power, which fluctuates in less regular patterns than \ac{PV}, even small model differences have a much stronger impact on optimal investments, for example availability assumptions in DIETER, and a different interpretation of specific investment costs in REMix. Therefore, small differences in modeling can substantially impact the optimal use and investment decisions of short-duration (battery) storage.

For long-duration (cavern) storage, deviations on a regional level are higher than for short-duration (battery) storage (Figure \ref{fig:diff_carvern_regions}). One reason are higher total investment costs for this technology that is mainly expanded to shift large amounts of energy from summer to winter (Figure~\ref{fig:storLevel2}). The differences between models are particularly pronounced in Belgium, France, Luxembourg, Poland, and Switzerland with deviations up to almost 100 \%. In Belgium, Luxembourg, and Switzerland the gap between base and maximum peak demand is small compared to other regions. At the same time, expanded capacities for long-duration (cavern) storage are comparatively small with respect to maximum peak loads. Consequently, we conclude that there is potentially less demand for long-duration storage in those regions having the effect that, because of small modeling differences, E2M2 and GENESYS-2 decide against expanding. The other models (DIETER, oemof, and REMix), on the contrary, decide to expand which leads to a high deviation between the two groups of models. However, in France the difference between base and maximum peak demand is the largest of all regions, and the summer-demand-share is lowest of all regions (Figure~\ref{fig:summer-demand-ratio}), which leads to higher demands for long-duration storage expansion. However, the implementation of a fixed \ac{E2P} ratio and a charge-to-discharge ratio for long-duration storage in E2M2 increases the costs for expanding this technology. For this reason E2M2 builds 96 \% less capacity compared to the average of all other models that invest heavily in this technology. The same applies to Poland with the difference that the main cause is the comparatively high expansion of long-duration storage towards base demand.

\subsection{Use case III: Expansion of thermal power plants and storage technologies}\label{ssec:expThermalAndStorage}

\begin{figure}[htp]
    \centering
    \includegraphics[width=\textwidth]{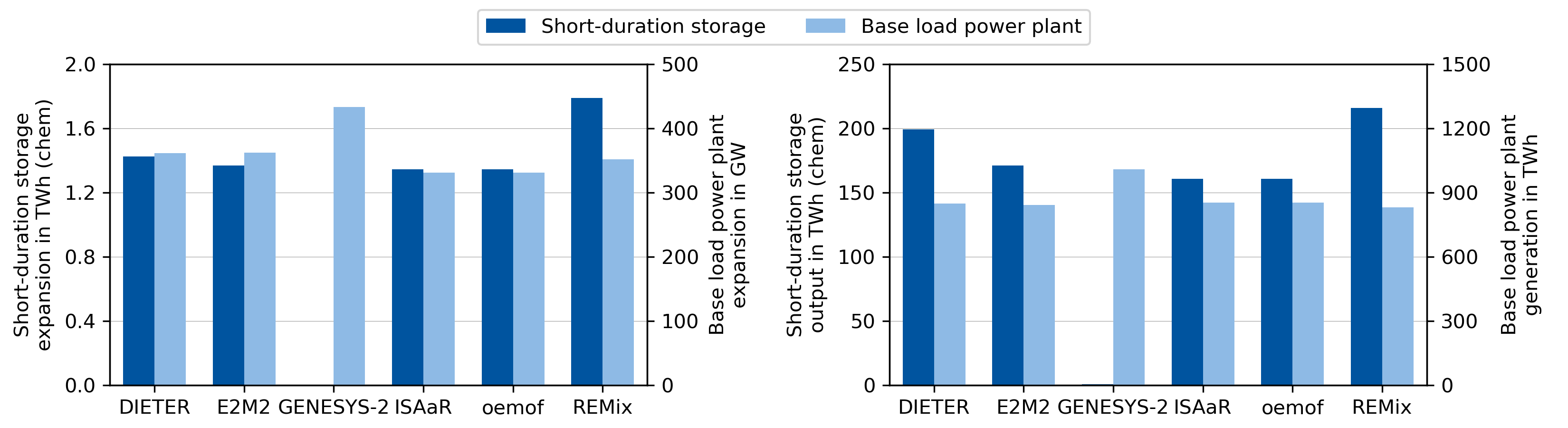}
    \caption{Capacity expansion and generation of short-duration (battery) storage and base load power plants across all regions.}
    \label{fig:capExpBat2c}
\end{figure}

Figure \ref{fig:capExpBat2c} shows optimal energy capacity of short-duration (battery) storage and generation capacity of base load (nuclear) power plants. Additionally, we show the discharge of short-duration storage and generation of base load power plants. 

The optimal capacity results highlight that DIETER and E2M2, as well as ISAaR and oemof, show very similar expansion. Comparing the two groups of results, the capacity expansion in ISAaR and oemof is 9\% lower for base load (nuclear) power plants. This difference occurs because DIETER and E2M2, in contrast to ISAaR and oemof, consider limited availability factors (Section \ref{ssec:expThermal}). Therefore, less expansion is required in models with full availability of components. For the expansion of short-duration (battery) storage the same conclusion applies. However, the limited availability of short-duration storage is at 98 \% such that the differences between the models are only minor.

Comparing REMix to the other models reveals substantially higher expansion of short-duration (battery) storage. As described in Section \ref{ssec:expStorage}, reduced specific costs of storage in REMix cause higher investments. At the same time, this makes it possible to reduce the use of fuel for base load power plants, and leads to lower system costs. 

Standing out from all models, GENESYS-2 only invests into base load power plants, and no short-duration (battery) storage capacity is built. This is because of a missing foresight window of the pre-defined dispatch approach. If installed, the available energy in the short-duration (battery) storage unit would not always suffice to meet the hourly peak demand in every time step (in combination with the base load generation capacity), because the decisions to fill up the storage would have to be made in past time steps. Instead, the optimal base load generation technology, which is assumed to be fully available at any given time, increases to be able to meet peak residual load. We conclude that models with a pre-determined dispatch order per design are not suitable for simplified use cases focusing on choices between a fully dispatchable generation technology and electricity storage. However, such a limited technology portfolio is unlikely to be applied beyond the model comparison conducted here. 

\subsection{Use case IV: Expansion of transmission capacities}\label{ssec:expTransmission}

\begin{figure}[htp]
    \centering
    \includegraphics[width=\textwidth]{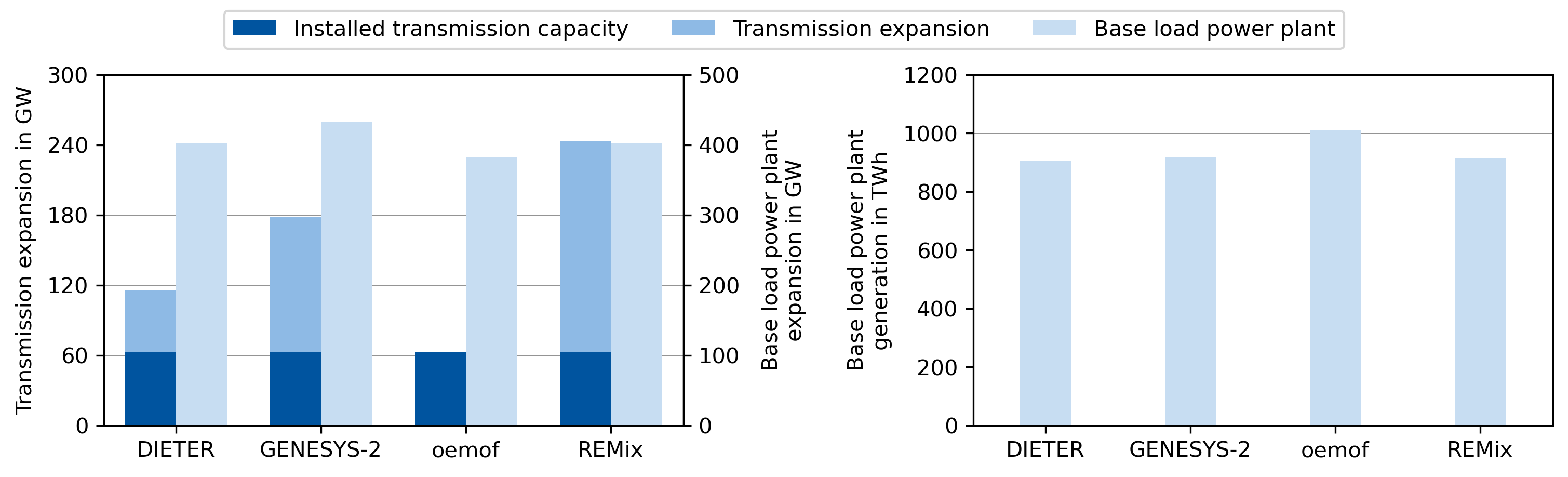}
    \caption{Capacity expansion of transmission lines and capacity expansion and generation of base load power plants across all regions.}
    \label{fig:capTransmission2d}
\end{figure}

Figure \ref{fig:capTransmission2d} illustrates the optimal expansion of transmission capacities and base load (nuclear) power plants. The existing transmission capacity sums up to 63.205 GW across all lines. The model oemof does not allow for transmission expansion. Its underlying transmission grid is limited to the exogenous endowment.

The expansion results show that REMix determines the highest optimal transmission capacity expansion, followed by GENESYS-2 and DIETER. The deviation between DIETER and GENESYS-2 is driven by differences in underlying modeling approaches. In GENESYS-2, the dispatch model favours a more regional use of energy. Domestic load is balanced with neighboring regions first. Only if this is not possible, balancing with more distant regions becomes available. In contrast, DIETER allows for spatial energy exchange without any regional preferences. Grid use across all regions is more evenly and efficiently distributed, which reduces investment cost into transmission infrastructure compared to GENESYS-2 (Figure~\ref{fig:costComparison}). However, the less efficient use of transmission capacity in GENESYS-2 leads to higher investment in base load (nuclear) power plants and generation from such in comparison with all other models. Consequently, the high investment cost for base load power plants are the true reason for the increased expansion cost in GENESYS-2.

Standing out from the other models, REMix uses a \ac{DCLF} grid representation, also accounting for the impact of regional grid use on the entire grid. A margin of the available transfer capacity of one line is used due to flows on other lines. Consequently, optimal transfer capacities increase in comparison with GENESYS-2 and DIETER. This shows, that in this stylized setting with a very limited set of available flexibility options, the simplified \ac{NTC} approach is likely to underestimate the true need for grid infrastructure. Despite this, the investment cost in REMix are lower than in GENESYS-2, and on the same level as DIETER (Figure~\ref{fig:costComparison}). Like in GENESYS-2, high costs for expansion of base load (nuclear) capacity, which are very similar in REMix and DIETER, make up the largest share of total investment cost such that differences in transmission expansion have only a minor effect.

Without the capability of expanding transmission capacities, an expansion of base load (nuclear) capacity is the only option to cover demand in oemof. Nevertheless, oemof shows lower expansion in comparison with DIETER and REMix, because those two models account for reduced availability of base load power plants (91.2~\%) (Section~\ref{ssec:expThermal}). This way, oemof requires less base load capacity and can generate more electricity from one unit than the other models. Considering the relatively high investment cost for base load power plants compared to transmission capacity, it becomes clear why oemof has the lowest overall investment cost in this use case (Figure \ref{fig:costComparison}).

\begin{figure}[htp]
    \centering
    \includegraphics[width=12cm]{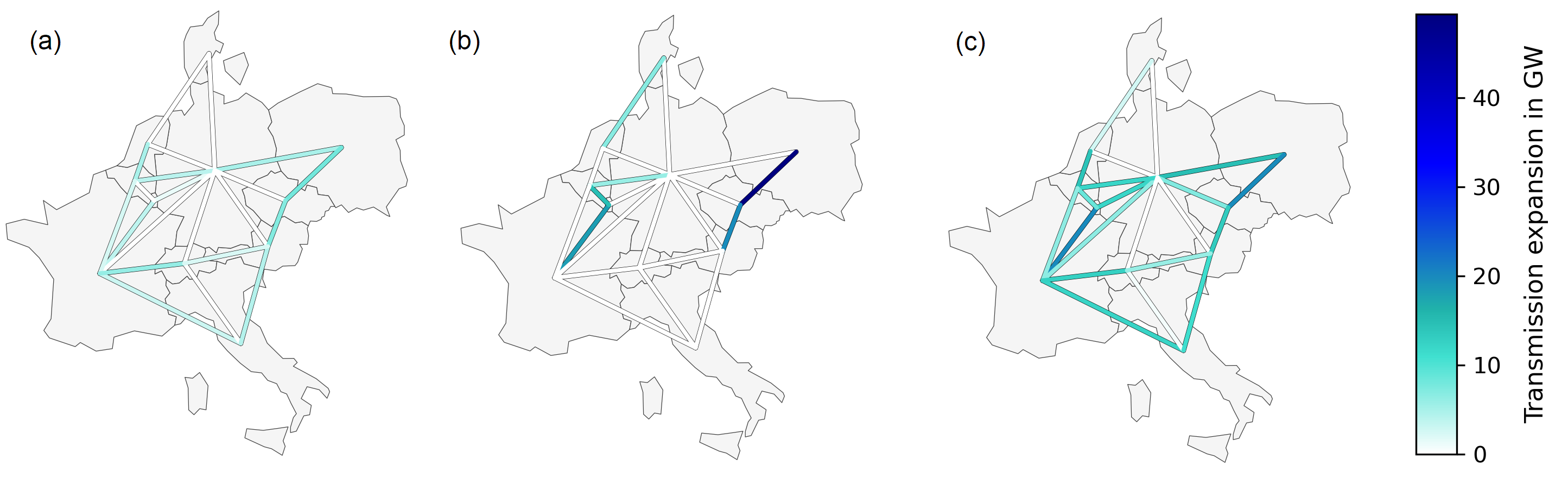}
    \caption{Geographical distribution of capacity expansion of transmission lines in (a) DIETER, (b) GENESYS-2 and (c) REMix}
    \label{fig:capTransmission2dgeo}
\end{figure}

On a region level, the results also differ because of different transmission modeling approaches. This is illustrated in Figure~\ref{fig:capTransmission2dgeo} with the geographical distribution of transmission expansion.

DIETER and REMix find a relatively similar optimal distribution of transmission expansion, yet with a level shift driven by the difference between \ac{NTC} and \ac{DCLF} representations. In contrast, GENESYS-2 only expands selected regional lines. This effect result from the dispatch hierarchy in GENESYS-2 that favours exchange of energy with regional neighbours rather than with distant regions. 

Focusing on the details, all models show a higher expansion from Poland and France to one or more neighbouring countries. In Poland, comparably high available wind generation (Figure~\ref{fig:capDiffRegion}) is one factor that increases possibility of exports. In addition, Poland only connects to two other countries which aggregates necessary grid exchange to only two transmission lines. Those characteristics cause a strong capacity expansion between Poland and Czech Republic of more than 40 GW in GENESYS-2. France combines a high wind generation availability with a relatively low summer-demand-share (Figure~\ref{fig:capDiffRegion}). Regions including high available \ac{PV} generation and a higher summer-demand-share, like Austria, Belgium, and Switzerland (Figure \ref{fig:capDiffRegion}), can be a supplement to France by balancing out generation and demand between the regions. This is one factor why we observe stronger grid expansion between France and such countries.

\section{Summary and conclusion}\label{sec:conclusion}

We compare six capacity expansion models of the power sector, drawing on four simplified use cases. These allow separating the effects from expanding different technologies by covering individual building blocks of a future energy system. In all use cases we identify deviations in expansion results and link them to modeling differences, taking into account overlapping effects. A comparison of expansion cost shows that deviations are highest when combining expansion of storage and base load power plants.

One modeling difference that occurs in all use cases is related to diverging assumptions on the availability of technologies. Modeling limited availability always causes higher investments to compensate for reduced availability of affected technologies. Our analysis in use case I further reveals that the effects of limited availability can be far more pronounced for peak load power plants than for base load power plants. Other factors that have a larger influence on expansion results are exogenous assumptions regarding the \ac{E2P} ratio and the charge-to-discharge ratio of storage technologies, which we analyze in use case II. When these ratios are not optimally predefined, we observe significant deviations in results as the costs for storage expansion increase. This leads to different ratios between expansion of competing short- and long-duration storage technologies. Apart from differences in storage dimensioning, we find that exogenous assumptions on initial and final storage levels can have an impact on the operation of long-duration storage.

Other deviations in capacity expansion result from differences in modeling approaches. Embedding a pre-ordered dispatch structure in a model generally leads to slightly lower capacity expansion of flexibility technologies like storage and transmission, but higher expansion of base or peak load power plants , due to reduced foresight and flexibility of energy supply. At the same time, this causes higher investment costs in comparison with \ac{LP}-models. In use case III, we show that models with a pre-ordered dispatch structure under rare circumstances can generate entirely different optimal expansion solutions. Nevertheless, we also conclude that these effects are much less pronounced in more detailed use cases with a larger and more realistic technology portfolio. 

Focusing on the expansion of transmission capacities, outcome deviations originate from fundamentally different modeling approaches. A simplified \ac{NTC} approach most likely underestimates the true need for grid infrastructure, in comparison with a more detailed \ac{DCLF} approach. Moreover, an analysis of the geographic distribution of transmission expansion reveals that models focusing on regional energy supply in a pre-ordered dispatch structure rather invest in single transmission lines, which increases expansion costs.

On a region level, we observe that for short-duration storage the biggest differences in results occur in regions with high a potential of wind power generation. In contrast, the highest deviations for long-duration storage are observed when the gap between base and peak is smallest, because the need for expensive long-duration storage decreases. Therefore, we generally conclude that between regions different levels of generation from \ac{VRE} and demand characteristics can increase the likelihood of deviations between model results. Modelers should generally be aware of this dependency. 

Despite the simplicity of use cases we expect that the general effects identified in our stylized setting also hold in more detailed model applications, although they may be less visible there. However, it will then be more difficult to isolate the effects, as they overlap or interfere with each other. Additionally, increasing model complexity related to sector coupling potentially complicates the analysis and comparison of different model results. To isolate effects from different sector coupling modeling approaches, we propose to apply our simplified approach in respective future model comparisons. Furthermore, the effect of regional characteristics on expansion results should be investigated further, as our findings highlights that complex correlations between generation and demand profiles are difficult to separate.

\section*{Acknowledgements}
The research for this paper was performed within the FlexMex project supported by the German Federal Ministry of Economic Affairs and Energy under grant number 03ET4077. The authors thank Georgios Savvidis, Felix Böing, Kai Hufendiek, Benjamin Fleischer, Christoph Weber, Berit Müller, Mascha Richter, Frank Merten, Dirk-Uwe Sauer, Niklas van Bracht, Arne Pöstges, and Caroline Podewski for contributing to the project conceptualization and funding acquisition.

\section*{Author contribution}
\textbf{Jonas van Ouwerkerk:} Conceptualization, Methodology, Software, Validation, Formal analysis, Investigation, Writing - Original draft preparation, Writing - Review \& Editing, Visualization
\textbf{Hans-Christian Gils:} Methodology, Software, Validation, Formal analysis, Investigation, Data curation, Writing - Original draft preparation, Writing - Review \& Editing, Visualization, Supervision, Project administration, Funding acquisition
\textbf{Hedda Gardian:} Methodology, Software, Validation, Formal analysis, Investigation, Data curation, Writing - Original draft preparation, Visualization
\textbf{Martin Kittel:} Methodology, Software, Validation, Formal analysis, Investigation, Writing - Original draft preparation, Writing - Review \& Editing
\textbf{Wolf-Peter Schill:} Investigation, Writing - Review \& Editing, Funding acquisition
\textbf{Alexander Zerrahn:} Methodology, Software, Validation, Formal analysis, Investigation
\textbf{Alexander Murmann:} Methodology, Software, Validation, Formal analysis, Investigation, Writing - Original draft preparation, Writing - Review \& Editing
\textbf{Jann Launer:} Methodology, Software, Validation, Formal analysis, Investigation,  Writing - Original draft preparation, Writing - Review \& Editing
\textbf{Laura Torralba-Díaz:} Methodology, Software, Validation, Formal analysis, Investigation, Writing - Original draft preparation, Writing - Review \& Editing
\textbf{Christian Bußar:} Writing - Review \& Editing, Software, Funding acquisition 

\section*{Data Availability}
The data template and input data used are available on [Link to be added].

\bibliography{bibfile}

\end{document}